\documentclass[preprint,amsmath,amssymb, aps,pre,floatfix,superscriptaddress,longbibliography]{revtex4-2}
\usepackage{graphicx} 
\usepackage[separate-uncertainty=true]{siunitx}
\usepackage[version=4]{mhchem}
\usepackage{comment}
\usepackage{xcolor}
\usepackage{tikz}
\usepackage{scalerel}
\usepackage{xifthen}
\usepackage{chngcntr}
\usepackage{xcolor}
\usetikzlibrary{svg.path}
\usepackage[colorlinks,allcolors=blue]{hyperref}

\newcommand{\changed}[1]{#1}
\newcommand{\figref}[2][{}]{Fig.\ \ref{#2}\ifthenelse{\isempty{#1}}{}{\,(#1)}}

\newcommand{\secref}[1]{Sec.\ \ref{#1}}

\newcommand{\plb}[1]{(\MakeLowercase{#1})} 
\renewcommand{\vec}{\mathbf}

\definecolor{orcidlogocol}{HTML}{A6CE39}
\tikzset{
  orcidlogo/.pic={
    \fill[orcidlogocol] svg{M256,128c0,70.7-57.3,128-128,128C57.3,256,0,198.7,0,128C0,57.3,57.3,0,128,0C198.7,0,256,57.3,256,128z};
    \fill[white] svg{M86.3,186.2H70.9V79.1h15.4v48.4V186.2z}
                 svg{M108.9,79.1h41.6c39.6,0,57,28.3,57,53.6c0,27.5-21.5,53.6-56.8,53.6h-41.8V79.1z M124.3,172.4h24.5c34.9,0,42.9-26.5,42.9-39.7c0-21.5-13.7-39.7-43.7-39.7h-23.7V172.4z}
                 svg{M88.7,56.8c0,5.5-4.5,10.1-10.1,10.1c-5.6,0-10.1-4.6-10.1-10.1c0-5.6,4.5-10.1,10.1-10.1C84.2,46.7,88.7,51.3,88.7,56.8z};
  }
}

\newcommand\orcid[1]{\href{https://orcid.org/#1}{\mbox{\scalerel*{
\begin{tikzpicture}[yscale=-1,transform shape]
\pic{orcidlogo};
\end{tikzpicture}
}{|}}}}

\newcommand\Pen{\mbox{\textit{Pe}}}

\begin{document}

\setlength{\unitlength}{1cm}

\newcommand{\goeaffila}{Max Planck Institute for Dynamics and Self-Organization, Am Fa\ss{}berg 17, 37077 G\"ottingen, Germany}
\newcommand{\goeaffilb}{Institute for the Dynamics of Complex Systems, Georg August Universit\"at G\"ottingen, Germany}

\newcommand{\goeaffil}{Max Planck Institute for Dynamics and Self-Organization, Am Fa\ss{}berg 17, 37077 G\"ottingen, Germany and Institute for the Dynamics of Complex Systems, Georg August Universit\"at G\"ottingen, Germany}

\newcommand{\twaffil}{Physics of Fluids Group, Max Planck Center for Complex Fluid Dynamics, and J.M.\ Burgers Center for Fluid Dynamics, University of Twente, PO Box 217,7500AE Enschede, Netherlands}
\newcommand{\hydaffil}{Department of Mechanical and Aerospace Engineering, Indian Institute of Technology Hyderabad, Kandi, Sangareddy, Telengana- 502285, India}
\newcommand{\pennaffil}{Dpt. of Physics \& Astronomy, University of Pennsylvania, 209 S 33rd St., Philadelphia, PA 19104, USA}

\newcommand{\princaffil}{Dpt. of Chemical and Biological Engineering, Princeton University, Princeton, NJ 08544, USA}

\newcommand{\grenaffil}{Universit\'e Grenobles-Alpes}
\newcommand{\yorkaffil}{Mathematics Department, University of York, Heslington, York, YO10 5DD, UK}

\title{Interfacial activity dynamics of confined active droplets}

 \author{Prashanth Ramesh~\orcid{0000-0003-3264-5084 }}
\affiliation{\goeaffila}
\affiliation{\twaffil}
\author{Babak Vajdi Hokmabad~\orcid{0000-0001-5075-6357}}
\affiliation{\goeaffila}
\affiliation{\princaffil}
\author{Arnold J.\ T.\ M.\ Mathijssen~\orcid{0000-0002-9577-8928}}
\affiliation{\pennaffil}
\author{Dmitri O.\ Pushkin~\orcid{0000-0001-5215-1618}}
\affiliation{\yorkaffil}
 \author{Corinna C.\ Maass~\orcid{0000-0001-6287-4107}}
 \email{c.c.maass@utwente.nl}%
\affiliation{\goeaffila}
\affiliation{\twaffil}
\date{\today}%

\begin{abstract}
%
{\changed Active emulsions can spontaneously form self-propelled droplets or phoretic micropumps. 
It has been predicted that the interaction with their self-generated chemical fields can lead to multistable higher-order flows and chemodynamic phenomena. 
However, it remains unclear how such reaction-advection-diffusion instabilities can emerge from the interplay between chemical reactions and interfacial hydrodynamics.
Here, we simultaneously measure the flow fields and the chemical concentration fields using dual-channel microscopy for oil droplets that dynamically solubilize in a supramicellar aqueous surfactant solution.
We developed an experimentally tractable setup with micropumps, droplets that are pinned between the top and bottom surfaces of a microfluidic reservoir, which we compare directly to predictions from a Brinkman squirmer model to account for the confinement.
With increasing droplet radius, we observe (i) a migration of vortex flows from the posterior to the anterior of the droplet, analogous to a transition from pusher- to puller-type swimmers, (ii) a bistability between dipolar and quadrupolar flow modes, and, eventually, (iii) a transition to multipolar modes.
We also investigate how the dynamics evolve over long time periods.
Together, our observations suggest that a local build-up of chemical products leads to a saturation of the surface, which controls the propulsion mechanism.
These multistable dynamics can be explained by the competing time scales of slow micellar diffusion governing the chemical buildup and faster molecular diffusion powering the underlying transport mechanism. 
Our results are directly relevant to phoretic micropumps, but also shed light on the interfacial activity dynamics of self-propelled droplets and other active emulsion systems.}
\end{abstract}

\maketitle
\section{Introduction}

Phoretic mechanisms are widely used to actuate transport, be it externally mediated, e.g. by chemo- or electrophoresis, or self-generated:
{\changed At macroscopic scales, phoresis can drive living organisms and culinary Marangoni cocktails \citep{mathijssen2022culinary}, and in microfluidic environments it can power} self-propelled microswimmers like active colloids or droplets \citep{izri2014_self-propulsion,herminghaus2014_interfacial,maass2016_swimming,babu2022_motile,birrer2022_we,dwivedi2022_self-propelled,michelin2023_self-propulsion}, or self-driven micropumps {\changed for nutrient or fuel mixing and advection in artificial and biological systems}~\citep{yu2020_microchannels,gilpin2017_vortex,antunes2022_pumping}. In self-actuated systems, fluid transport is induced through non-equilibrium processes at the interface between the particle and the surrounding fluid where a slip velocity and/or tangential stresses are generated \citep{anderson1989_colloid}. Various phoretic mechanisms have been reported based on the type of interactions between the particle and the ambient fluid. Regardless of the type of activity, all mechanisms rely on the inhomogeneity of a surrounding field such as gradients in solute concentration \citep{golestanian2005_propulsion}, temperature \citep{young1959_motion}, and electric fields \citep{bazant2004_induced-charge}.

In the specific case of a self-driven chemically active particle or droplet, the net motion is achieved by converting chemical free energy released by the agent into mechanical work through two distinct mechanisms: the diffusiophoretic effect, introducing a  finite slip velocity at the agent interface, and the Marangoni effect  creating interfacial tangential stresses.  {\changed In this context, we define \textit{activity} as the conversion of undirected, chemical kinetics into mesoscopic advective flow, either driving pumping motion or a translation of the particle itself.}
\par
Reactivity gradients can be implemented by design, i.e. a  built-in asymmetry, for instance by variation of the coating thickness for Janus particles \citep{ebbens2014_electrokinetic,campbell2019_experimental}, in binary systems of interacting particle pairs \citep{reigh2015_catalytic,reigh2018_diffusiophoretically}, or in droplets with adsorbed colloidal caps~\citep{ikcheon2021_interfacially-adsorbed}.
\par
\begin{figure}
\noindent
\centering
\includegraphics[width=1\textwidth]{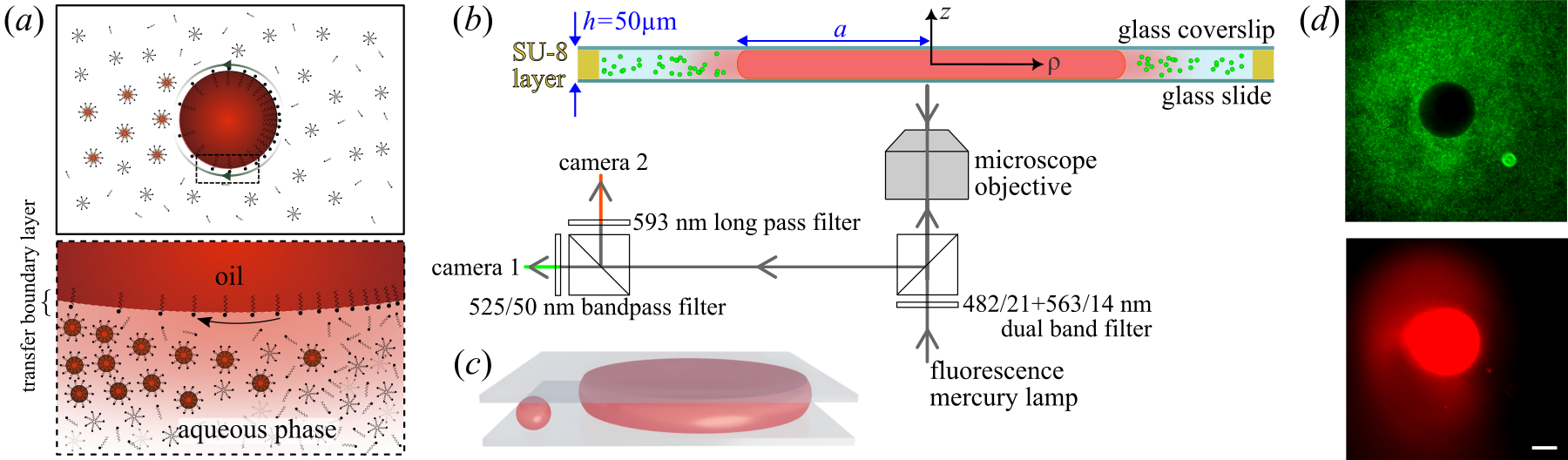}
\caption{\plb{a} Schematic of droplet propulsion mechanism and a zoomed in region around the droplet interface showing surfactant monomers, empty micelles and oil-filled micelles. \plb{b} Schematic of the dual-channel microscopy setup simultaneously visualising flow fields (camera 1) using fluorescent tracers and filled micelle concentration (camera 2) using Nile Red dye in the oil droplet. {\changed  \plb{c} Illustration of a small, spherical self-propelling droplet vs. a larger droplet pinned by squeezing between the glass top and bottom of the cell. \plb{d} Emission from fluorescent tracers (top) and Nile Red (bottom). Scale bar: \SI{50}{\um}}. }
\label{fig:1}
\end{figure}

However, starting from a spherically isotropic particle, the only viable route to generate an interfacial reactivity gradient is by interaction with the chemical products. 
{\changed Chemically active droplets are one of the prime model systems studied in this context. Here, an advection-diffusion instability in the chemical field around the droplet interface causes interfacial flow modes, ranging, with increasing Péclet number \Pen, from an inactive isotropic base state over a dipolar state that is propulsive in non-pinned droplets, to higher order extensile modes \citep{michelin2013_spontaneous}.
\par This emergence of higher interfacial modes and increasingly complex mesoscopic droplet motion with increasing \Pen\ has been found to be quite general in experiments, irrespective of whether \Pen\ is varied by way of chemical activity, droplet size or viscosity~\citep{izzet2020_tunable,suda2021_straight-to-curvilinear,hokmabad2022_spontaneously,hokmabad2021_emergence}.
\par
It was shown in \cite{michelin2013_spontaneous} and subsequent works that the most general prerequisite to set this advection-diffusion activity into motion is simply the consumption of a single chemical species at a constant rate at the interface. However, experimental realisations typically require more complicated chemical pathways. A popular model system consists of oil-in-water (alternatively, water-in-oil) emulsions in which the outer phase contains fuel supplied by a surfactant at supramicellar concentrations~\citep{herminghaus2014_interfacial,izri2014_self-propulsion,izzet2020_tunable,meredith2020_predator-prey,suda2021_straight-to-curvilinear}.
Here, the oil phase continuously solubilizes by migrating from the bulk droplet into empty or partially saturated surfactant micelles, leading to an interfacial Marangoni gradient that powers a self-supporting active interface (\figref{fig:1}a). \par
Surfactant micelles and monomers constitute distinct species that diffuse at significantly different timescales, and these complex kinetics have been found to affect the details of the observed dynamics.

Experimental examples include} the  emergence of bimodal motility through micellar accumulation at hydrodynamic stagnation points~\citep{hokmabad2021_emergence}, or trail avoidance stemming from chemotactic repulsion by long-lived chemical traces of other droplets~\citep{jin2017_chemotaxis,moerman2017_solute-mediated,lippera2020_alignment,lippera2020_collisions,izzet2020_tunable,hokmabad2022_chemotactic}. 
{\changed Current analytical and numerical work is looking beyond the one-species approximation, e.g. considering inhibitory effects of oil-filled micelles \citep{morozov2020_adsorption}, necessitating coupled advection-diffusion equations for species (large micelles vs. monomers) with different individual diffusion timescales and \Pen. At the current state of the art, such studies are somewhat hampered by best guesses at currently unknown molecular dynamics like aggregation kinetics, as well as computational constraints, that e.g. limit micellar aggregation numbers below typical values from experimental literature. Thus, while we see intriguing predictions, e.g. regarding multi-stable states (coexisting different vortex patterns), there is no one-to-one comparison between theory and experiment yet.

\par 
We therefore require well-controlled experimental paradigms to suggest testable geometries and provide realistic feedback to the assumptions of theory, 

We propose such a system in the present study on pinned droplets, or micropumps. Here, one does not have to account for the displacements caused by the often chaotic mesoscopic motion of a motile system.  Specifically, pinning gives access to steady state experimental conditions that allow for precise, simultaneous measurements of hydrodynamic and chemical fields.

In the experiments presented below, we measure the flow and chemical fields generated by pinned active droplets, and fit the vortex structure in the bulk medium to an analytical model of flow in a Brinkman medium, which allows for a qualitative analysis of the interfacial modes. We explore \Pen\ space via changing the droplet radius,  analyse steady state and long-term time dependent flow patterns around these pinned droplets with respect to anterior/posterior symmetry and multistable states and posit hypotheses on how these patterns are shaped by the multispecies chemodynamics of the droplet solubilisation. 

}

\par

\section{Methods} 
\subsection{Materials and experimental protocol}
\par
Our experimental system consists of (S)-4-Cyano-4'-(2-methylbutyl)biphenyl  (CB15, Synthon Chemicals) oil droplets immersed in an aqueous solution of the cationic surfactant tetradecyltrimethylammonium bromide (TTAB, Sigma-Aldrich). {\changed We note that while we have used the nematogen 5CB in some of our previous studies \citep{kruger2016_curling,kruger2016_dimensionality}, here we are using its branched isomer CB15 (or 5*CB) which is, while being a widely used chiral dopant, isotropic in its pure form at room temperature~\citep{mayer1999_polymorphism}.} CB15 droplets of different sizes were obtained by shaking. The TTAB surfactant concentration was fixed at 5wt.\% (CMC = 0.13 wt.\%). Our experimental reservoir consisted of {\changed a SU-8 photoresist spacer layer spin coated on glass, framing a  rectangular cell of area $13\times\SI{8}{mm^2}$ and height of \SI{50}{\um} fabricated via UV photolithography. We filled the cell with a dilute droplet emulsion and sealed it with a glass cover slip.} In our analysis we included droplets with typical radii larger than \SI{50}{\um}, which were compressed into flat disks of radius $a\gtrsim \SI{60}{\um}$. These disks were pinned at the top and bottom of the reservoir and exhibited only pumping motion. {\changed This pinning is only effective at relatively low surfactant concentration (here, 5wt\%) -- and might either be caused by oil/water/glass contact line pinning, or by short range no-slip interactions in a boundary layer of surfactant and water molecules between oil and the oleophobic glass surface. The droplet shape we observed is a flat cylinder with a convex oil-water interface, as sketched in Fig~\ref{fig:1}c. }

\subsection{Double-channel fluorescent microscopy}\label{sisec:DCFM}
To simultaneously image chemical and hydrodynamic fields, we have adapted an Olympus IX-83 microscope for dual-channel fluorescent microscopy, as shown in the light path schematic in \figref{fig:1}b. The fluorophores were excited using a fluorescence mercury lamp which is passed through a dual band excitation filter 482/21+563/14\,\SI{}{\nm}. The oil phase was doped with the co-moving Nile Red (Sigma-Aldrich) dye to label oil-filled filled micelles, with an emission peak of \SI{630}{\nm}. The tracer particles (FluoSpheres, yellow-green fluorescent, \SI{0.5}{\um} in diameter),  visualising the fluid flow around the droplet, emit light at a maximum of $\sim \SI{510}{\nm}$. The emission was separated into two channels using a beam splitter and appropriate filters for each emission maximum, and recorded by two 4MP cameras (FLIR Grasshopper 3, GS3-U3-41C6M-C), at 24\,fps for the green (tracers) and the red channel (dye) for short time measurements. Figure \ref{fig:1}d shows emission from fluorescent tracers (top) and Nile red (bottom). For long time measurements, we recorded the slowly changing chemical field continuously at a reduced frame rate of 4\,fps. To record tracer colloids, a higher frame rate of 24\,fps was required. Due to data storage limitations, we did not record continuously for the entire experiment, but recorded \SI{5}{\s} sequences at \SI{30}{\s}  intervals. Using  10X and 20X objectives focused on the cell mid-plane, the region of observation spanned \SI{1113}{\um^2} and \SI{557}{\um^2}, {\changed with a focal depth of \SI{3.06}{\um} and \SI{1.10}{\um}, respectively. We therefore assume all extracted flow data to represent the mid-plane values in $x,y$ to good accuracy.}
\subsection{Image processing and data analysis}\label{sisec:improc}

We extracted the droplet position and radius $a$ in ImageJ using the time averaged signal from the colloidal tracer channel, where the pinned droplet appears as a dark area, $A$.
{\changed We estimate the errors in extracted contour roundness and radius, $\Delta a/a$ to be $<1\%$.}

We extracted the flow field around the droplet by particle image velocimetry (PIV) of the tracer particles using the MATLAB-based  PIVlab interface~\citep{thielicke2014_pivlab}, with the droplet area $A$ used as a mask. For the PIV analysis, we chose a $48\times48$ pixel interrogation window with $67\%$ overlap, at a spatial resolution of \SI{0.28}{\um}/px. We then fit the 2D flow data generated by this procedure to the a hydrodynamic model described below in \secref{sec:brinkman} {\changed within the microscope's field of view, using a least-squares fit algorithm. As PIV is unreliable near boundaries, we exclude an expanded area around the droplet, using $\rho>1.1a$.}

For further analysis we required time dependent chemical and tangential flow fields around the droplet perimeter $a\theta$ with the polar angle $\theta$ taken counter-clockwise from the droplet anterior (see dashed circles in Figure~\ref{fig:5}a,b). The flow data was taken from PIV as close to the droplet as possible, i.e. at  $\rho=1.1a$. For the chemical field, we used the channel recording the Nile red fluorescence and extracted the intensity from an annular region around $A$. Since the chemical signal was oversaturated  around $A$ due to the high concentration of dye in the droplet itself, we collected  fluorescence data at a distance from the interface being $\rho=1.22a$. {\changed We separately investigated  the effects of varying the droplet size on the short time steady state flow (section~\ref{sec:3_1}), as well as time-dependent saturation effects over long-time measurements on the order of \SI{20}{\minute} (section~\ref{sec:3_4}). During these long-time measurements the droplet radius reduced by approximately 7\% due to solubilisation.} We made sure to only include data from the first 1--\SI{2}{\minute} of experiments in the investigation of size effects shown in section~\ref{sec:3_1}.

\subsection{Theory: Brinkman squirmer in 2D}\label{sec:brinkman}

To model the flow in the aqueous phase, we consider a {\changed cylindrical} squirmer {\changed of radius $a$} at low Reynolds number confined between two parallel plates separated by height $h$. 
Because confinement strongly affects the flows generated by microswimmers \citep{mathijssen2016_hydrodynamics, jeanneret2019_confinement,mondal2021_strong}, we follow the framework given by \cite{jin2021_collective} to calculate the flow field $\vec{u}(x,y)$ in the mid plane between the plates. Specifically, we approximate the 3D Stokes equations by the 2D Brinkman equations  \citep{brinkman1947_calculation,tsay1991_viscous,pepper2010_nearby,gallaire2014_marangoni}:
    \begin{equation}
        \vec{\nabla} p = \mu (\nabla^2 - k) \vec{u}, \quad \vec{\nabla} \cdot \vec{u} = 0,
    \end{equation}
Here the permeability is defined as $k = 12/h^2 = 1/\lambda^2$, in direct analogy with Darcy's law \citep{whitaker1986_flow}, and $\lambda$ is the slip length.

For a pinned droplet, the oil-water interface corresponds to a circle around the coordinate origin with radius $\rho = a$ in polar coordinates $(\rho,\theta)$. 
We assume that this interface is impermeable and that the flow field is determined by the tangential velocity, here given by a harmonic expansion using the Bessel functions of the second kind $K_n$:
\begin{align}u_{\rho}^n(\rho=a)&=0 &\text{and} && u_{\theta}^n(\rho=a)&=-\dfrac{a}{\lambda}\dfrac{K_{n-1}(a/\lambda)}{K_{n}(a/\lambda)}b_n \sin(n\theta). \end{align}
The flow field $\vec{u}$ in the 2D domain  around the droplet, {\changed  corresponding, in the Hele-Shaw cell, to the $z$ averaged flow in ($x,y$),} is then given by the pumping solution:
\begin{equation}
    {u_{\rho}}^{n}=b_n \left[ n \left(\frac{a}{\rho}\right)^{n+1} -n\frac{a}{\rho}\frac{K_n(\rho/\lambda)}{K_n(a/\lambda)} \right]\cos(n\theta)
\end{equation}
\begin{equation}
    {u_{\theta}}^{n}=b_n \left[ n \left(\frac{a}{\rho}\right)^{n+1} +\frac{a}{\lambda}\frac{K'_n(\rho/\lambda)}{K_n(a/\lambda)} \right]\sin(n\theta)
\end{equation}

The flow patterns around the droplet are characterised by the first two coefficients $b_1,b_2$: First, their weights determine the dominant flow mode, which is dipolar for $|b_2/b_1|<1$ and quadrupolar for $|b_2/b_1|>1$ \citep{nganguia2018_squirming,jin2021_collective}. 
{\changed For illustration, the supplementary figure S8 shows the calculated stream lines for increasing $|b_2/b_1|$ in the Brinkman model.}
Second, for a primarily dipolar flow pattern, the sign of $b_2$ determines the location of the dominant vortex pair: $b_2>0$ results in an anterior and $b_2<0$ a posterior location, with the droplet anterior defined at $\theta=0$. In our fits of the experimental velocity field to this model, we use $b_1,b_2$ as fit parameters with $b_n=0$ for $n>2$, and, figure~\ref{fig:6} excepted, all $xy$ data plots have been rotated to have the droplet's anterior or $\theta=0$ axis to point to the right, in positive $x$. 

{\changed We note that, using the Brinkman description, we assume that $u(z)\approx0$: all monopolar flow modes like a gravitational force monopole (cf.  \cite{deblois2019_flow,hokmabad2022_spontaneously} for sedimenting droplets) should be suppressed by the strong $z$ confinement of the Hele-Shaw geometry.}

\section{Results}\label{sec:results}

\begin{figure}
\centering
\includegraphics[width=1\textwidth]{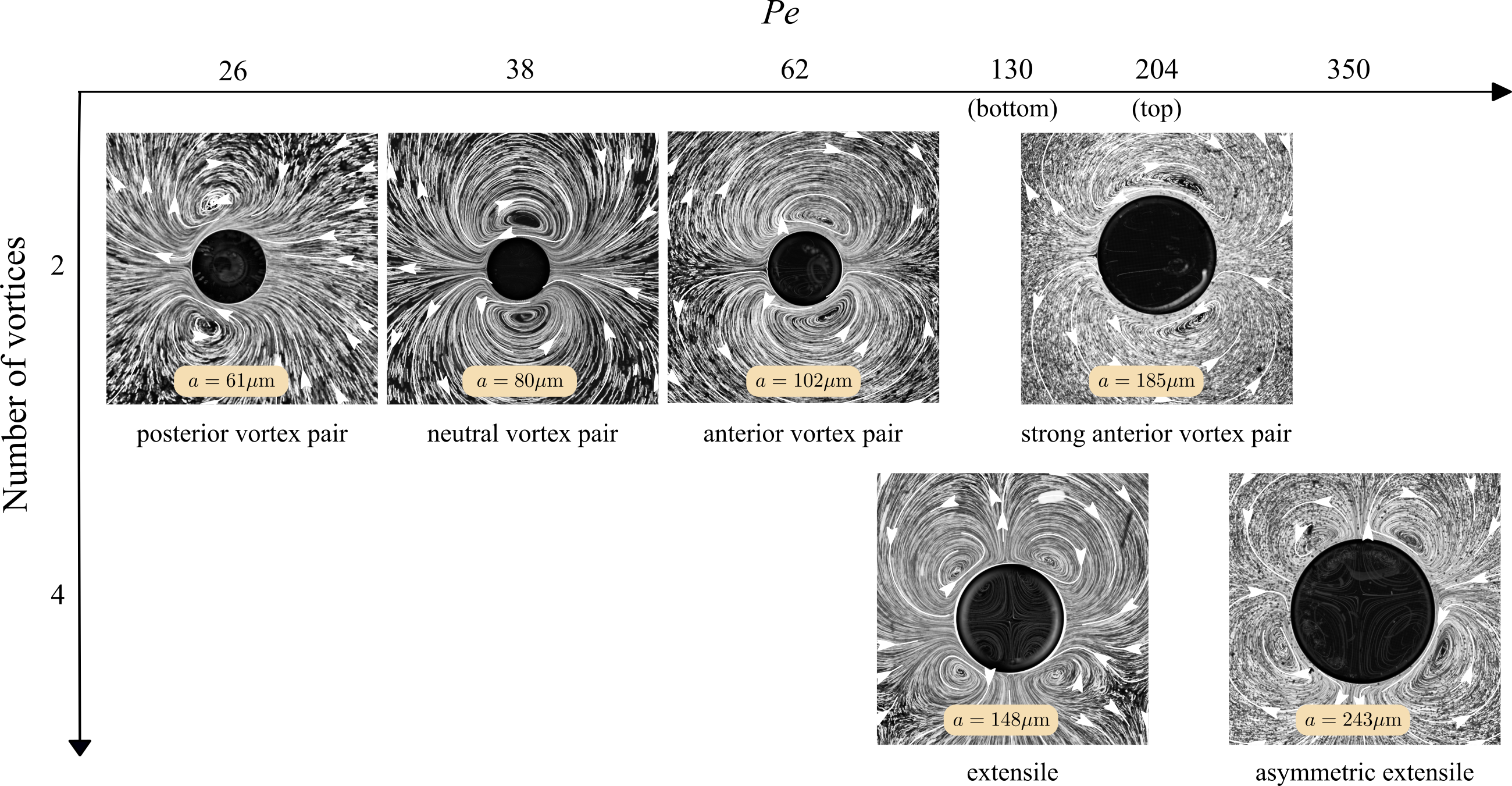}
\caption{Streak visualisation~\citep{gilpin2017_flowtrace} of flow fields generated by pumping active droplets for increasing droplet radius $a$, with superimposed streamlines from PIV analysis. We observe a shift of the vortex pair from the droplet posterior to the anterior in the top row, corresponding to $a=\SI{61}{\um}$, $a=\SI{80}{\um}$, and $a=\SI{102}{\um}$, and on further increase of $a$, a bistable regime between a dipolar ($a=\SI{185}{\um}$) and a quadrupolar flow mode ($a=\SI{148}{\um}$). For larger radii, we only see the quadrupolar mode ($a=\SI{243}{\um}$). See also videos S1 and S2.}
\label{fig:2}
\end{figure}

For droplets driven by  micellar solubilisation, self-sustaining propulsive or pumping flows develop  when the Péclet number $\Pen$ exceeds a critical threshold \citep{michelin2013_spontaneous}. $\Pen$ is given by the ratio of advective to diffusive transport of surfactant monomers from the droplet interface to oil-filling micelles~\citep{herminghaus2014_interfacial}, and
 can be estimated {\changed in the single-species picture} as follows (cf. eqn. 1 and appendix B.2 in~\cite{hokmabad2021_emergence}, following \cite{anderson1989_colloid,izri2014_self-propulsion,morozov2019_nonlinear}):
\begin{equation}
\Pen\approx \frac{18 \pi^2}{k_BT}  q_s r_s^2 \zeta a^2 \mu^i \left[\mu \left(\frac{2\mu+3 \zeta/a}{2 \mu+3}\right)\right]
\end{equation}
Here, $a$ is the droplet radius, $\mu=\mu^o/\mu^i$ the viscosity ratio between outer and inner medium, $\zeta \sim \SI{10}{\nm}$ the characteristic length scale over which surfactant monomers interact with the droplet,   $r_s \sim \SI{10e-10}{\m}$ the length of a monomer, and 
$q_s = D \textbf{n} \cdot \nabla c$ is the isotropic interfacial surfactant desorption rate per area. 
\par
{\changed We use this definition of \Pen\ for the present geometry as well, since the underlying chemical processes are identical: an increase in the droplet radius $a$ corresponds to an increase of $\Pen$.}

\subsection{Flow field characterisation and comparison with the Brinkman squirmer model}\label{sec:3_1}

\par
We begin with a classification of the observed flow patterns around pumping flat droplets, as shown in figure \ref{fig:2}, and the supporting videos S1 and S2. For small radii, $a=\SI{61}{\um}$, or $\Pen=26$, we find one posterior droplet pair. With increasing droplet radius, $a=\SI{80}{\um}$ or $\Pen=38$, the vortex pair shifts towards the midline (`neutral' position) and finally to the droplet anterior, for $a=\SI{102}{\um}$ with $\Pen=62$. \par
On further increasing the droplet radius, we observe a bistable regime, featuring either a vortex pair displaced further towards the droplet anterior ($a=\SI{185}{\um}$, $\Pen=204$) or  a symmetric extensile or quadrupolar flow field ($a=\SI{148}{\um}$, $\Pen=130$). 
For even bigger droplet radii we find only the quadrupolar mode, with an increasing asymmetry in the vortex positions which shift towards the anterior ($a=\SI{243}{\um}$, $\Pen=350$).
\par
\begin{figure}
\centering
\includegraphics[width=1\textwidth]{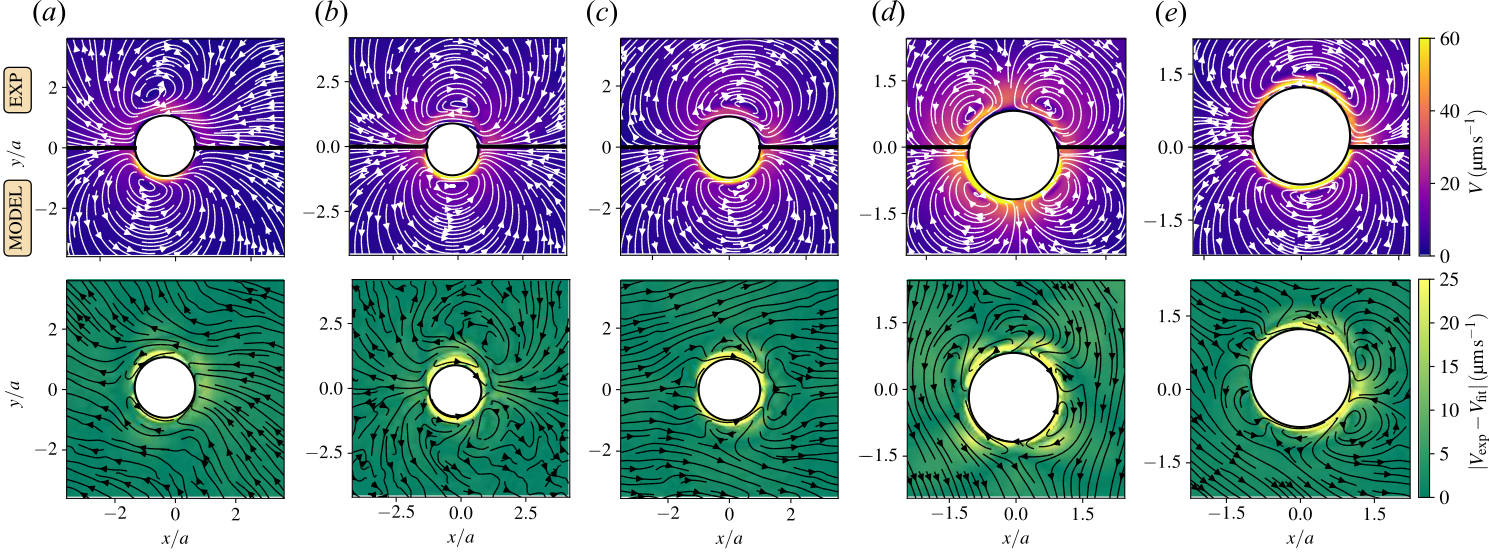}
\caption{Top row: Comparison between the experimental flow fields $v_\text{exp}$ (top half of each panel) and the Brinkman model fits  $v_\text{fit}$ (bottom half) for different $\Pen$, with fit parameters $b_1,b_2$. \plb{a} $\Pen=26$ ($b_2/b_1=-0.529$), \plb{b} $\Pen=38$ ($b_2/b_1=-0.001$), \plb{c} $\Pen=62$ ($b_2/b_1=0.191$), \plb{d} $\Pen=130$ ($b_2/b_1=4.915$) and \plb{e} $\Pen=204$ ($b_2/b_1=0.175$). Depending on the $b_2/b_1$ ratio, we note a shift of the vortex pair towards the droplet anterior, and a bistability of dipolar and quadrupolar patterns for large droplets. Bottom row: Residual flow fields, | $v_\text{exp}- v_\text{fit}|$.}
\label{fig:3}
\end{figure}

 
For a quantitative {\changed evaluation of the interfacial modes}, we measure the flow fields by PIV and compare them to the Brinkman model by fitting the flow fields via  $b_1,b_2$ as shown in figure \ref{fig:3} where the panels (a)--(e) analyse the data presented in figure \ref{fig:2}. The streamlines in the top row show a good quantitative agreement between experimental data (top half) and the Brinkman model (bottom half). 

 {\changed The model successfully captures the vortex positions in the bulk medium, as shown by the comparison of measured and fitted streamlines in the top row; and by the small and unstructured residual flow fields in the far field (bottom row).}

In analogy to the squirmer model for motile droplets \citep{ishikawa2006_hydrodynamic,downton2009_simulation}, we can define a parameter $\beta=b_2/b_1$ that indicates the vortex centre position. A posterior vortex pair corresponds to $b_2/b_1<0$, neutral vortices to $b_2/b_1=0$ and anterior vortices to $b_2/b_1>0$. A quadrupolar configuration is caused by a dominant second mode with $|b_2/b_1|>1$. However, for dipolar anterior vortices (figure \ref{fig:3}e), the fit to the Brinkman model {\changed up to second order somewhat}  under-predicts $\theta_v$, which leads to distinct vortex structures in the residual field.


\subsection{Squirmer parameter and phase diagram}\label{sec:3_2}

\begin{figure}
\centering
\includegraphics[width=1\textwidth]{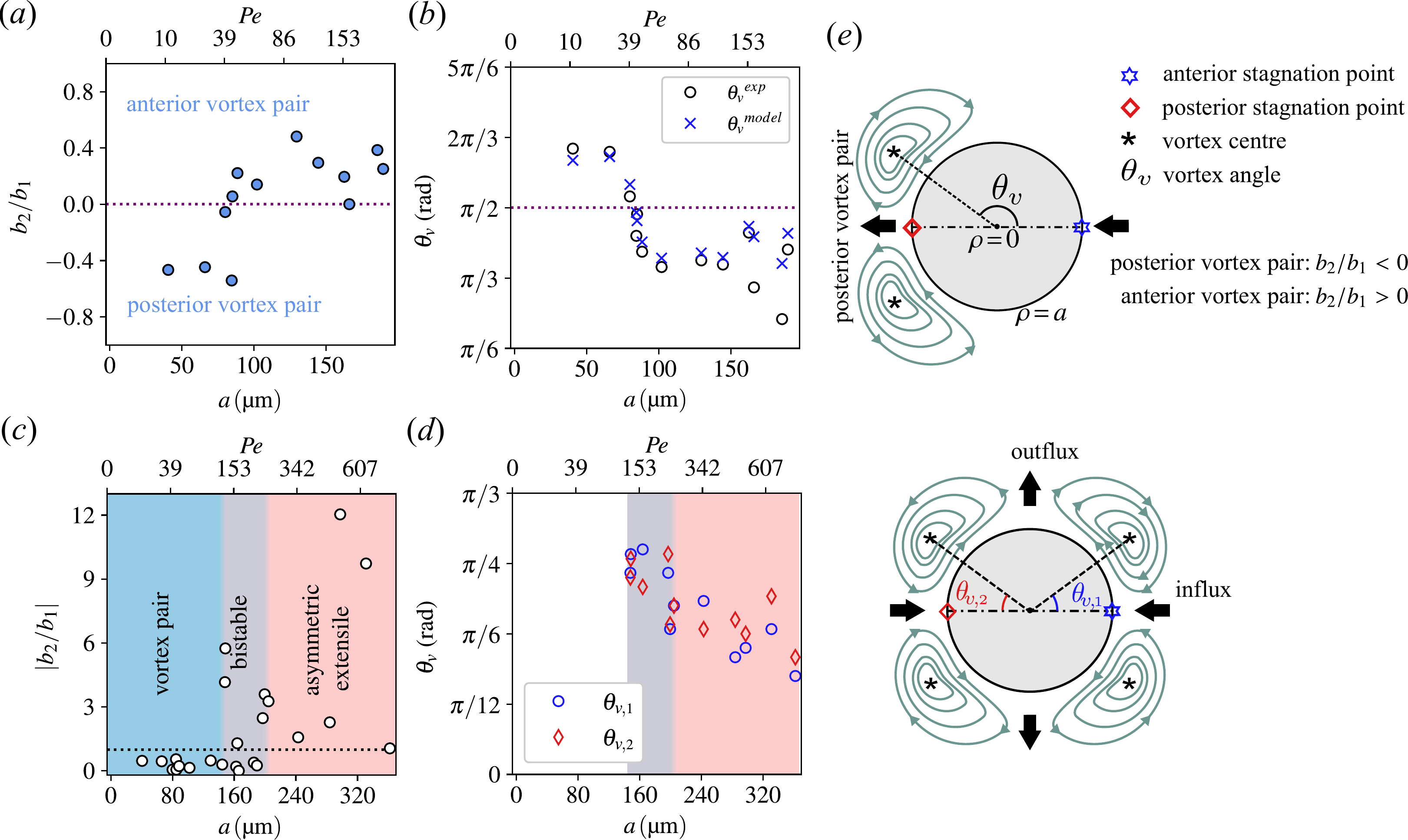}
\caption{\plb{a,c} Variations of squirmer parameter $b_2/b_1$ versus $\Pen$. \plb{b,d} Variations of angle between front stagnation point and vortex centre $\theta_v$ versus $\Pen$. \plb{e} Schematic defining the vortex angle.}
\label{fig:4}
\end{figure}

\par
Using the data obtained by fitting the Brinkman model to the experimental data, we plot the change of the squirmer parameter $b_2/b_1$ with the  droplet radius $a$ in figure \ref{fig:4}a. With increasing radius, $b_2/b_1$ increases from negative to zero and then to positive values, but still remains smaller than one, which indicates a dominant first mode. This $b_2/b_1$ transition reflects the vortex shift from the droplet posterior to a neutral and then an anterior position. We further calculate the vortex angle $\theta_v$, defined as the polar angle of the line between $\rho=0$ and the {\changed location of} $\vec{u}(x,y)=0$ at the centre of the respective vortex (figure \ref{fig:4}e). To compare the vortex position predictions from the Brinkman model, we plot the vortex angles obtained from both experiment and theory versus the radius in figure \ref{fig:4}b. We observe an excellent agreement up to large radii of $a\gtrsim \SI{150}{\um}$. Here, with a strong anterior shift, the experimental vortex centre position is underpredicted by the model.
\par
At larger radii, we predominantly find  quadrupolar flow fields, \textit{i.e.} a dominant second mode. For classification, we plot the different modes in the space spanned by $|b_2/b_1|$ and $a$ as shown in figure \ref{fig:4}c. The phase diagram is divided into three regions, namely, \textit{(i)} `vortex pair', \textit{(ii)} `bistable', featuring either a strong anterior vortex pair or a symmetric extensile flow and \textit{(iii)} `asymmetric extensile', where the anterior vortex pairs are displaced inwards. The dashed horizontal line, $|b_2/b_1|=1$, separates dipolar (below) and quadrupolar (above) flow fields. For large droplets, the quadrupolar state also loses fourfold symmetry, as the vortices move towards the anterior and posterior stagnation points. To characterise this asymmetry, we plot the vortex angles for both anterior, $\theta_{v1}$, and posterior, $\theta_{v2}$, vortex pairs in figure \ref{fig:4}d. For the symmetric quadrupolar state, the vortex angle is $\theta_v\approx\pi/4$ (grey region), whereas in the asymmetric configuration, $\theta_v < \pi/4$ (pink region). We note that for both dipolar and quadrupolar asymmetric states, the vortices are pulled towards stagnation points with radial influx.

\subsection{Simultaneous chemical and flow field characterisation}\label{sec:3_3}
\par
{\changed For a deeper analysis, we simultaneously measure the droplet flow fields and the chemical concentration fields using dual channel microscopy as described above in section~\ref{sisec:DCFM}. Previous studies have shown that oil-filled micelles are considerably larger than the reactive species, \textit{i.e.} surfactant monomers and will therefore diffuse much more slowly \citep{hokmabad2021_emergence}, such that we can expect them to aggregate at stagnation points and cause secondary inhibition effects  as proposed in~\cite{morozov2020_adsorption}}.
Here, oil-filled micelles are labelled by Nile red dye co-moving with the oil phase.
\par
\begin{figure}
\centering
\includegraphics[width=1\textwidth]{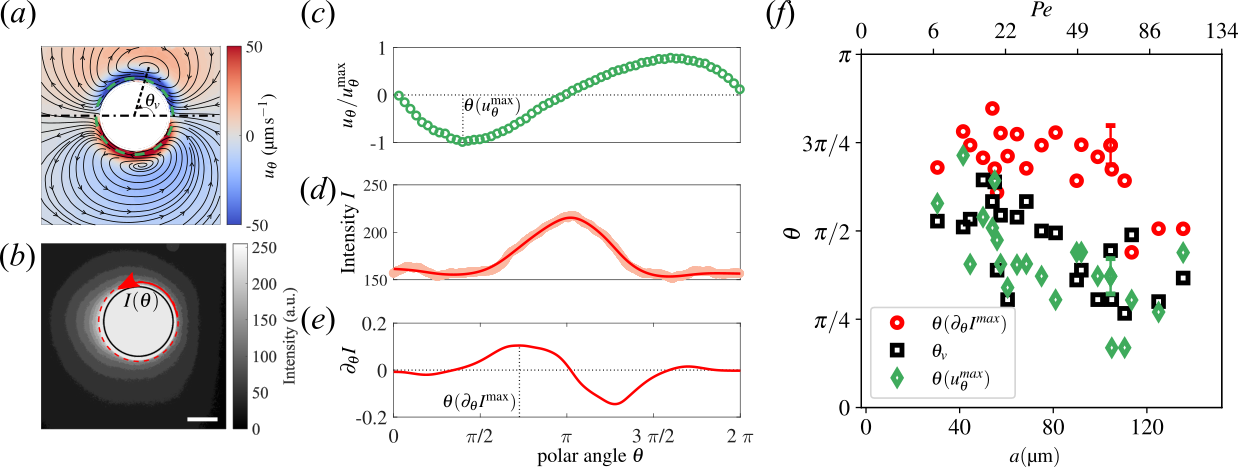}
\caption{(a) Flow streamlines around a pumping droplet at $\Pen=65$ colour coded by tangential velocity. Tangential velocity $u_{\theta}$ around the droplet interface is extracted along the green dashed circle. (b) Corresponding contour plot of fluorescent emission by Nile red doped oil droplet and filled micelles. Intensity profile around the droplet interface is extracted along the red circle. Scale bar: \SI{50}{\um}. (c) Tangential velocity profile at the interface. (d) Intensity profile of filled micelle concentration around the droplet. (e) corresponding gradient profile. {\changed(f) Variations of vortex angle (black), maximum tangential velocity (green) and maximum chemical gradient location around the droplet (red) versus $\Pen$. We have added a representative error bar that quantifies the systematic error due to the measurement being away from the droplet interface. All three observables decrease with increasing $\Pen$.}}
\label{fig:5}
\end{figure}

In figure \ref{fig:5} (a-e), we set out the relation between chemical and flow field for a close to neutral dipolar droplet.
Panel (a) maps the field of the tangential velocity $u_\theta$, with superimposed streamlines, while panel (b) displays a contour plot of the Nile red emission $I$. We further extract the profiles $u_\theta$ and $I(\theta)$ around the droplet along the dashed lines, as displayed in panels (c) and (d), with the gradient of the intensity profile in panel (e).

If we now assume that the interfacial surfactant density is depleted in the presence of filled micelles, an extremum of the gradient in the chemical field $\partial_\theta I^\text{max}$ should correlate to a Marangoni gradient in the interface and therefore to an extremum $u_\theta^\text{max}$ of the tangential velocity. Furthermore, the Brinkman model solution predicts that the polar angle $\theta_v$ for the stagnation points of the vortex pairs coincides with the angle of maximum tangential flow $\theta(u_\theta^\text{max})$. We have plotted the respective angles of $\theta_v$, $\theta(u_\theta^\text{max})$ and $\theta(\partial_\theta I^\text{max})$ in panel (f). 

As the droplet radius $a$ increases, the vortex position shifts from the posterior at $\theta_v>\pi/2$ to the anterior at $\theta_v<\pi/2$. In good agreement with the Brinkman model, $\theta(u_{\theta}^\text{max})$ coincides with $\theta_v$. $\theta(\partial_\theta I^\text{max})$ shows a similar shift with increasing droplet radius, but with an offset towards higher $\theta$ values compared to $\theta_v$ and $\theta(u_\theta^\text{max})$. This may be due to a systematic overestimation -- as noted in section~\ref{sisec:improc}, the fluorescence has to be measured away from the interface, where the micelles have already been advected further towards the droplet posterior.

This displacement of maximal chemical gradient location with increasing droplet radius suggests a saturation effect by oil-filled micelles that determines the vortex position $\theta_v$, {\changed which we will discuss in detail below.}

\par
\subsection{Long time evolution of the pumping droplet}\label{sec:3_4}

\begin{figure}
\centering
\includegraphics[width=1\textwidth]{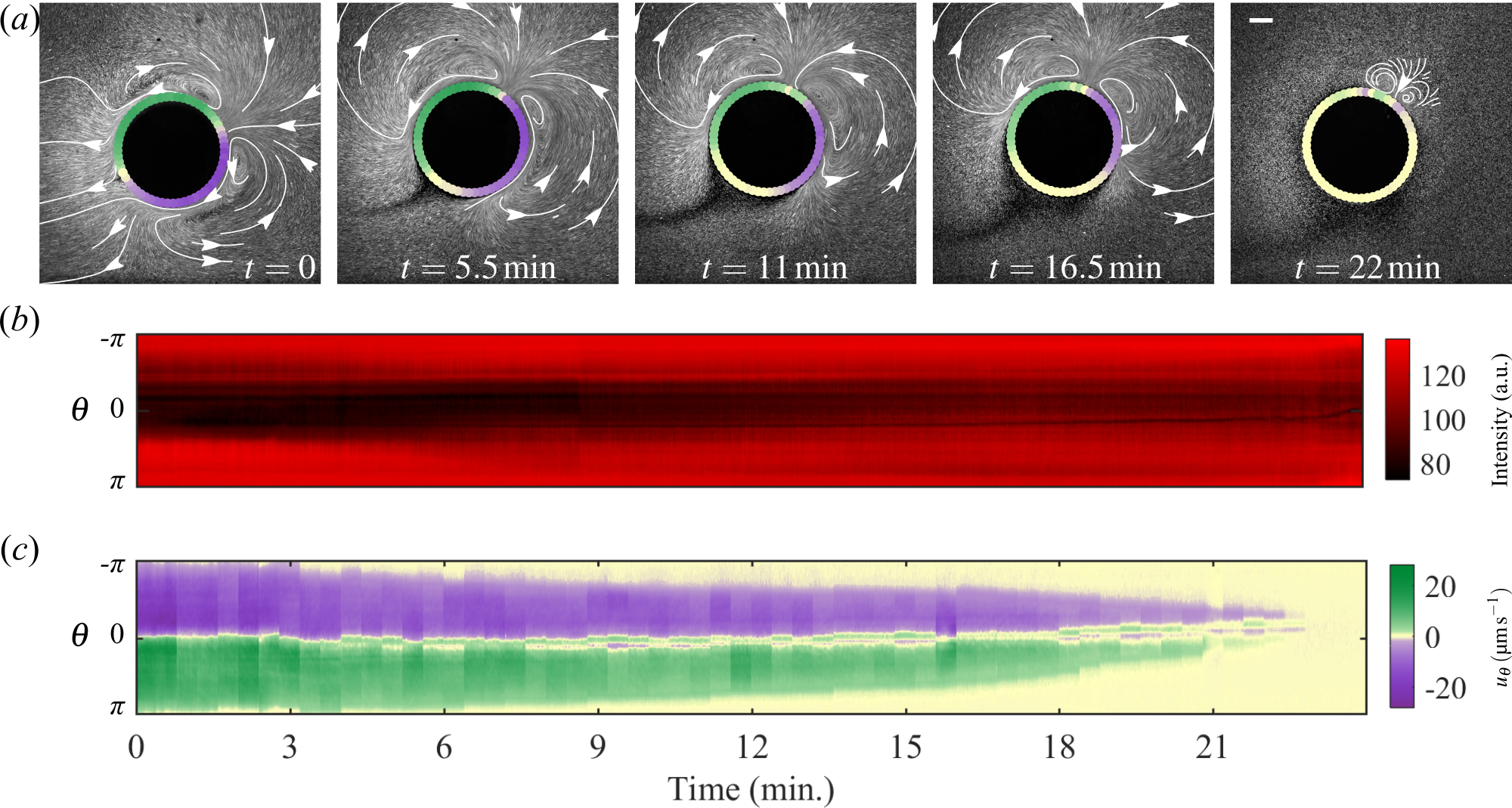}
\caption{\plb{a} Time evolution of streamlines around a pumping droplet at $\Pen=241$ and $a=\SI{201}{\um}$. As time increases, the inactive region starting from the rear stagnation point grows around the interface due to saturation by filled micelles. This leads to droplet inactivity at long time. Scale bar: \SI{100}{\um}. \plb{b} Time evolution of filled micelle concentration around the droplet interface. $\theta=0$ indicates the anterior stagnation point. \plb{c} Time evolution of tangential velocity around the droplet interface. {\changed See also Video S3.}}
\label{fig:6}
\end{figure}

{\changed Until now, we have looked at short-time dynamics. We will now investigate long-term dynamics, where self-interactions become more important: We expect reaction products to recirculate and interact with the droplet again after timescales longer than the advective one. This could influence the long-term interfacial activity dynamics.}
To quantify this, we recorded dual-channel microvideographs of flow and chemical fields around a droplet with a dipolar flow field, and analysed the long-time interfacial dynamics  (see also Video S3).
\par
Figure \ref{fig:6}a shows, for increasing time, colloidal tracer images with selected streamlines that illustrate the extracted flow field. We have plotted the tangential velocity at the interface, using the diverging colour map from panel (c) below, such that green corresponds to positive (counterclockwise) and purple to negative flow, while yellow regions mark negligible flow with $|u_{\theta}|<\SI{1}{\um\per\s}$, as expected for both the anterior and posterior stagnation points. At $t=0$, the entire remaining droplet perimeter is active.  With increasing time, as the droplet solubilises, oil-filled micelles build up and accumulate around the posterior stagnation point $\theta=\pm\pi$. In this area, the tangential flow speed decreases. The inactive region grows over time, as shown in the broadening yellow section of the tangential velocity map, while the vortices shift towards the anterior. 
 
At long times ($t=22$ min.), only a very small region at the droplet anterior is active and soon the entire droplet is rendered inactive.
\par
To track the continuous evolution of filled micelle concentration and tangential velocity at the droplet interface, we have mapped them onto kymographs in $t,\theta$ space in figure \ref{fig:6} (b) and (c). At $t=0$, the red fluorescence from oil filled micelles is centred around the posterior stagnation point at $\theta=\pm\pi$; the dark region around the anterior stagnation point indicates mostly empty micelles. With increasing time, as the number of oil-filled micelles increases, the red region at the posterior broadens until it fills almost the entire droplet perimeter. This is also evident from figure \ref{fig:6}c, where the green and purple bands of tangential flow corresponding to the two vortices narrow over time and recede towards $\theta=0$, while the yellow region broadens, indicating inactivity with $|u_{\theta}|<1\mu m/s$. 
For long times, this gradual saturation renders the droplet entirely inactive. {\changed This saturation is different from what to expect in motile droplets as they can escape their own chemical exhaust~\citep{hokmabad2021_emergence}.}
\par

\par
\subsection{The effect of micelle saturation}
{\changed Finally, we want to discuss two observations: First,} with increasing droplet size and $\Pen$, we have seen a displacement trend from posterior to anterior for both vortex-pair location, and maximum concentration gradient location. Second, long time measurements showed a saturation effect by filled micelles which leads to similar vortex location displacement, and eventually the stopping of activity.
{\changed This observation is different from numerical work on models assuming a constant interfacial activity by interaction with a single chemical species: a good example is provided in Fig.~3 of \cite{morozov2019_nonlinear}, where, with increasing $\Pen$, the dipolar vortex centre changes from midline symmetry, to a pusher-type posterior location, to a quadrupolar state with front-back symmetry.

We note that these models use a a spherical 3D boundary with an axisymmetric constraint and no external forces: we are not aware of studies on a Brinkman quasi-2D system model with a pinning force. 

We may however speculate that the different evolution of vortex patterns is mediated by the solubilization mechanism, to which a generalised one-species model is agnostic.

Micelles will not absorb an infinite number of oil molecules (else the solubilization would create new droplets, which defies thermodynamics). Assuming, to first order, a constant flux of absorbing micelles along an active interface, and a constant rate at which they absorb oil from the boundary layer, there should be a fixed saturation time $t_s$, and associated length beyond which micelles will not fill any further. This length scale limits the size of the vortex patterns around the droplet.}

{\changed To extract this length scale, we analyze the steady state flow fields around of droplets of different sizes (Section \ref{sec:3_1}) with respect to two quantities. First,  the length $a\theta_{v}$, which measures the arc length from the anterior stagnation point to the vortex centre. We assume $a\theta_{v}$ to scale with the saturation length, as $\theta_{v}$ marks the angle where the vortex streamlines begin to reorient away from the interface.} Second, the  average tangential velocity at the droplet interface $\overline{u}_\theta=\langle u_{\theta}(\rho=a,\theta )\rangle_{\theta=0.001}^{\theta=\theta_{v}}$,
which is calculated using the Brinkman model fit to the PIV flow field measurements. The saturation time $t_s$ defined above should now scale with $t_s=a\theta_{v}/\overline{u}_\theta$. 

\begin{figure}
\centering
\includegraphics[width=0.8\textwidth]{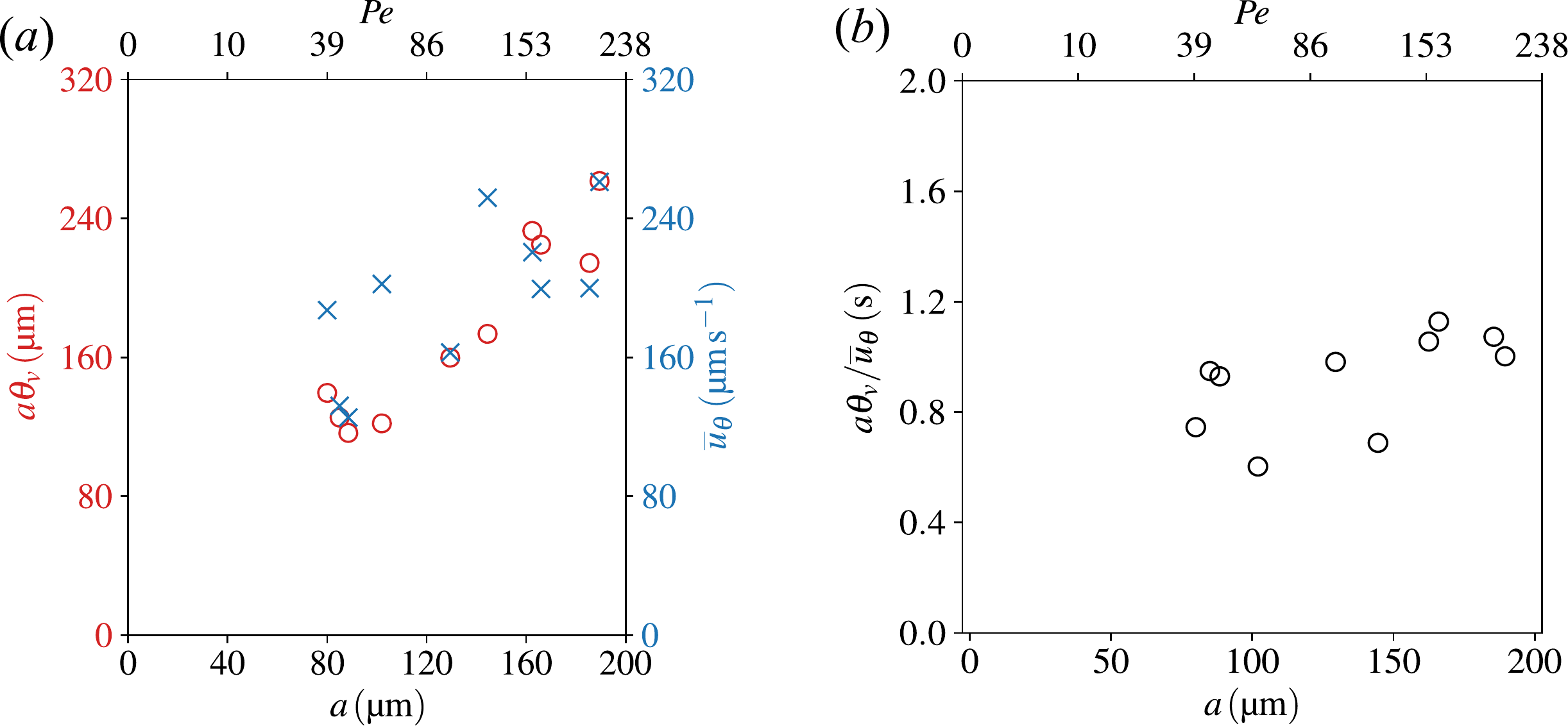}
\caption{{\changed (a) Vortex arc length $a\theta_{v}$ and averaged tangential velocity $\overline{u}_\theta$ versus droplet radius $a$. (b) Saturation time $a\theta_{v}/\overline{u}_\theta$ versus droplet radius $a$, as extracted from the quantities plotted in (a).} }
\label{fig:7}
\end{figure}

We have plotted the experimentally determined $a\theta_v$, $\overline{u}_\theta$ and $t_s$ vs. $a$ in figure~\ref{fig:7}. We find that both $a\theta_v$ and $\overline{u}_\theta$ increase with the droplet radius $a$ and  that $t_s=\SI{0.92(17)}{\s}$ is constant.  This indicates that for a fixed surfactant concentration the time for the micelles to become saturated is {\changed independent from the size of the droplet, and that this time scale sets the region of interfacial activity and thereby the anterior/posterior vortex configuration. We note that this argument does not hold for small droplets where the vortices are displaced beyond the posterior stagnation point, as $a$ needs to be  sufficiently large for the micelles to saturate while they are passing the interface. 

Given the above assumptions, it also follows  that for long times, i.e. beyond the time scale of the  steady state analysed above, the continuous recirculation of filled micelles advects fewer empty micelles at the droplet anterior, which fill up faster. This leads to a gradual decrease of the saturation times and length scales, which is what we observe in  figure~\ref{fig:6} and Video S3.}

\section{Discussion and outlook}
Chemically active droplets produce the flow fields that make them self-propelling microswimmers or pinned micropumps, primarily via Marangoni gradients caused by chemical reactions at the interface. In the experiments detailed above, we have studied chemical and flow fields around pinned droplets depending on both size and activity duration and performed a hydrodynamic mode analysis using a Brinkman squirmer model to account for quasi 2D confinement and a Péclet number $\Pen$ adapted to the specific chemical system. This allows us to compare our findings to recent developments in numerical and analytical models using similar $\Pen$ based mode stability approaches. 

We {\changed confirmed} that the extrema in interfacial flow, and thereby the structure of the vortex field around the droplet, are correlated to gradients in the local concentration of reactivity products, which shift around the perimeter with both increasing droplet size and progressing time. We have seen multistability of dipolar and quadrupolar modes, and an overall tendency of posterior-to-anterior displacement with increasing droplet size or experimental progress, which, {\changed during the observation of a droplet for a longer time}, leads eventually to an inactive droplet, where the interface is fully fouled up by chemical buildup. The analysis of our size-dependent data strongly suggests that the vortex structure is determined by the time scale of fuel conversion, as evidenced by a time independent micellar saturation time. 

Our findings provide illustration to recent developments in analytical advection-diffusion models. Historically, minimal models assumed an uniform interfacial reaction rate~\citep{michelin2013_spontaneous,izri2014_self-propulsion}, which indeed yields higher order propulsive and extensile (\textit{e.g.}, dipolar to quadrupolar) modes with increasing $\Pen$ with good semi-quantitative experimental agreement. However, open questions remain for further investigation. Notably, simulations~\citep{morozov2019_nonlinear} found a transition from neutral squirmer $\beta=0$ to pusher, $\beta<0$ with a posterior vortex pair, to quadrupolar extensile with increasing $\Pen$,  while our experiments presented above show an \textit{increase} of $\beta$ with increasing $\Pen$, {\changed albeit in a different, non-pinned, geometry}. {\changed However, i}n experiments on motile droplets driven by micellar solubilisation, sufficiently small swimmers are commonly weak pushers~\citep{deblois2019_flow,jin2021_collective,suda2021_straight-to-curvilinear}. Pusher-to-neutral tendencies, {\changed i.e. an increase of $\beta$}, with increasing droplet radius have also been reported for moving droplets~\citep{suda2021_straight-to-curvilinear}, suggesting that the divergent behaviour is  not a consequence of pinning.
Bistability of dipolar and quadrupolar states was found numerically for both a 3D swimmer held stationary~\citep{desai2021_instability}, as well as in a model where the interfacial reactivity was assumed to be inhibited by oil-filled micelles~\citep{morozov2020_adsorption}. Interestingly, in the latter case, the region of multistability increased with the inhibition parameter. 
The flow field of a Brinkman squirmer with a localised interfacial gradient, corresponding to our assumption of a predominantly inactivated interface for long times,  has been evaluated as well~\citep{gallaire2014_marangoni}, yielding very similar vortex patterns. We note, however, that these calculations were based on an preimposed interfacial tension gradient (or, experimentally, localised heating), such that the underlying physical mechanism is different from the one causing our self-evolving profile.

\par
Beyond the implications for individual motility, multistable and higher mode flow patterns matter in the collective dynamics of active matter, and in their interactions with confined geometries. \cite{tan2022_odd} showed that developing starfish embryos generate different flow patterns during its development process, which determine their self-organisation (formation, dynamics, and dissolution) into living crystals. 
Analogously, in artificial autophoretic systems, rotational instabilities on the individual scale~\citep{suda2021_straight-to-curvilinear,hokmabad2021_emergence} can carry through to the collective dynamics, where we have recently found $\Pen$ dependent stability and collective rotation in self-assembled planar clusters~\citep{hokmabad2022_spontaneously}. Here, the superposition of self-generated chemical fields might well provide a mechanism for the emergence of stable collective rotational states.
Our analysis of interfacial activity dynamics is also directly relevant to the study of active carpets \citep{mathijssen2018_nutrient, guzman-lastra2021_active}.

Impermeable {\changed boundaries like walls} will also affect the spreading of inhibitory chemical products and affect the stability of hydrodynamic modes, as recently demonstrated for a droplet near a wall~\citep{desai2021_instability}. Experimentally, such inhibitory effects in confinement can immobilise consecutive droplets in an active Bretherton scenario~\citep{deblois2021_swimming} or cause reorientation up to self-trapping~\citep{jin2017_chemotaxis,lippera2020_alignment,hokmabad2022_chemotactic}.

Understanding and modelling such phenomena from the molecular scale upwards is a daunting task, given the manifold species one has to keep track of and that the complex multistable dynamics are likely initiated by the competing time scales of slow micellar diffusion governing the chemical buildup and faster molecular diffusion powering the underlying advection-diffusion mechanism. {\changed We therefore propose the tools and experimental geometry in this study as a well-defined and quantifiable testing case to investigate in matching theoretical modelling. }

\section{Acknowledgments}
We gratefully acknowledge fruitful discussions with Prof. Jörn Dunkel at MIT and experimental optics support by Dr. Kristian Hantke at MPI-DS. We expressly acknowledge the contribution of experimental data by Myriam Rahalia, whom we were unable to contact at the time of submission.

%

\section{Supplementary material}
\subsection{Supplementary video captions}
 \textbf{Movie S1:} Flowtrace videos {\changed under bright-field microscopy} corresponding to three vortex patterns in Fig.2, with $a=\SI{61}{\um}$, $a=\SI{80}{\um}$, and $a=\SI{102}{\um}$ (left to right). The vortices around a pinned pumping droplet shift from the posterior to the anterior with increasing droplet diameter. Video clips are played in real time {\changed at 40fps} and repeated multiple times for ease of viewing. 
 
 \textbf{Movie S2:} Flowtrace videos {\changed under bright-field microscopy} corresponding to three vortex patterns in Fig.2, with $a=\SI{148}{\um}$, $a=\SI{185}{\um}$, and $a=\SI{243}{\um}$ (left to right). The vortices around a pinned pumping droplet are multistable, showing, with increasing size, symmetric quadrupolar, strongly asymmetric dipolar, and, for the largest droplet, asymmetric quadrupolar patterns. Video clips are played in real time {\changed at 40fps} and repeated multiple times for ease of viewing. {\changed Note that in the first panel, the oil droplet contained traces of \SI{1}{\um} diameter colloidal silica particles leading to serendipitous visualisation of droplet internal flow field as well.}
 
 \textbf{Movie S3:} Dual-channel videomicroscopy underlying the data in Fig. 6 (the interface of a droplet being gradually saturated by chemical buildup). Black/white frame and green channel track colloidal tracers (Flowtrace), the red channel the fluorescent emission from oil filled micelles with Nile Red marker. The droplet itself is masked in black. Time in experiment displayed in minutes.

\begin{figure}
    \centering
    \includegraphics[width=1\textwidth]{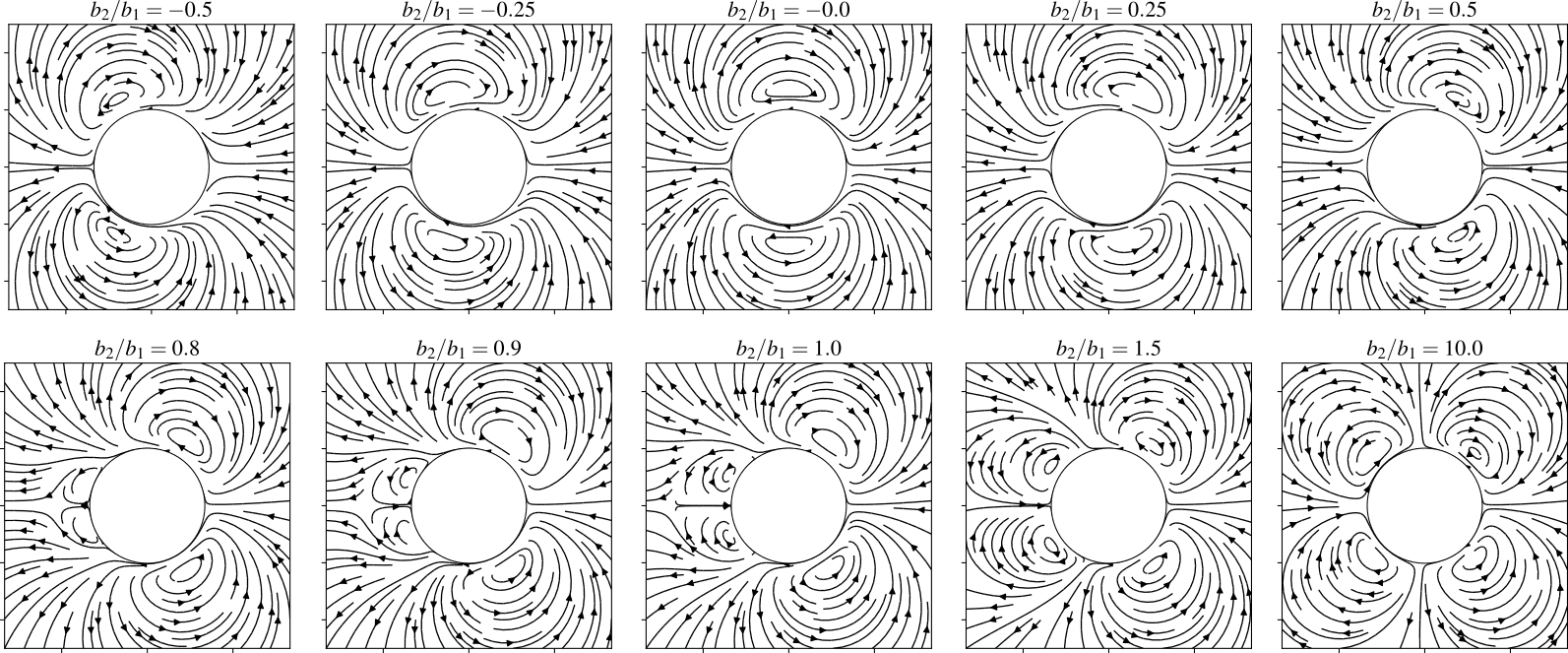}
    \caption{Pumping flow field around a 2D Brinkman squirmer for increasing ratio of the squirmer parameter $b_2/b_1$.}
    \label{fig:my_label}
\end{figure}

\end{document}